\documentclass[journal]{IEEEtran}
\usepackage{amssymb, amsmath}
\usepackage{float,epsfig}
\usepackage[all]{xy}
\usepackage{graphicx}
\usepackage{caption}
\usepackage{subcaption}
\usepackage{flushend}

\ifCLASSINFOpdf
\else
\fi
\begin{document}

\newtheorem{theorem}{\bf Theorem}[section]
\newtheorem{proposition}[theorem]{\bf Proposition}
\newtheorem{observation}[theorem]{\bf Observation}
\newtheorem{definition}[theorem]{\bf Definition}
\newtheorem{corollary}[theorem]{\bf Corollary}
\newtheorem{example}[theorem]{\bf Example}
\newtheorem{exam}[theorem]{\bf Example}
\newtheorem{remark}[theorem]{\bf Remark}
\newtheorem{lemma}[theorem]{\bf Lemma}
\newcommand{\nrm}[1]{|\!|\!| {#1} |\!|\!|}

\newcommand{\ba}{\begin{array}}
\newcommand{\ea}{\end{array}}
\newcommand{\von}{\vskip 1ex}
\newcommand{\vone}{\vskip 2ex}
\newcommand{\vtwo}{\vskip 4ex}
\newcommand{\dm}[1]{ {\displaystyle{#1} } }

\newcommand{\be}{\begin{equation}}
\newcommand{\ee}{\end{equation}}
\newcommand{\beano}{\begin{eqnarray*}}
\newcommand{\eeano}{\end{eqnarray*}}
\newcommand{\inp}[2]{\langle {#1} ,\,{#2} \rangle}
\def\bmatrix#1{\left[ \begin{matrix} #1 \end{matrix} \right]}
\def \noin{\noindent}
\newcommand{\evenindex}{\Pi_e}

%\newcommand {\proof} {\par{\it Proof}. \ignorespaces}
%\newcommand {\eproof}
%      {\space
%        {\ \vbox{\hrule\hbox{\vrule height1.3ex\hskip0.8ex\vrule}\hrule}}
%        \par}

%%%%%%%%%%%%%%%%%%%%%%%%%%%%%%%%%%%%%%%%%%%%%%%%%%%%%%%%%%%%%%%%%%%%%%%%%%

\def \R{{\mathbb R}}
\def \C{{\mathbb C}}
\def \K{{\mathbb K}}
\def \J{{\mathbb J}}
\def \Lb{\mathrm{L}}

\def \T{{\mathbb T}}
\def \Pb{\mathrm{P}}
\def \N{{\mathbb N}}
\def \Ib{\mathrm{I}}
\def \Ls{{\Lambda}_{m-1}}
\def \Gb{\mathrm{G}}
\def \Hb{\mathrm{H}}
\def \Lam{{\Lambda_{m}}}
\def \Qb{\mathrm{Q}}
\def \Rb{\mathrm{R}}
\def \Mb{\mathrm{M}}
\def \norm{\nrm{\cdot}\equiv \nrm{\cdot}}

\def \P{{\mathbb P}_m(\C^{n\times n})}
\def \A{{{\mathbb P}_1(\C^{n\times n})}}
\def \H{{\mathbb H}}
\def \L{{\mathbb L}}
\def \G{{\mathcal G}}
\def \S{{\mathbb S}}
\def \sigmin{\sigma_{\min}}
\def \elam{\sigma_{\epsilon}}
\def \slam{\sigma^{\S}_{\epsilon}}
\def \Ib{\mathrm{I}}
\def \Tb{\mathrm{T}}
\def \d{{\delta}}

\def \Lb{\mathrm{L}}
\def \N{{\mathbb N}}
\def \Ls{{\Lambda}_{m-1}}
\def \Gb{\mathrm{G}}
\def \Hb{\mathrm{H}}
\def \Delta{\triangle}
\def \Rar{\Rightarrow}
\def \p{{\mathsf{p}(\lam; v)}}

\def \D{{\mathbb D}}

\def \tr{\mathrm{Tr}}
\def \cond{\mathrm{cond}}
\def \lam{\lambda}
\def \sig{\sigma}
\def \sign{\mathrm{sign}}

\def \ep{\epsilon}
\def \diag{\mathrm{diag}}
\def \rev{\mathrm{rev}}
\def \vec{\mathrm{vec}}

\def \sk{\mathsf{skew}}
\def \sy{\mathsf{sym}}
\def \en{\mathrm{even}}
\def \odd{\mathrm{odd}}
\def \rank{\mathrm{rank}}
\def \pf{{\bf Proof: }}
\def \dist{\mathrm{dist}}
\def \rar{\rightarrow}

\def \rank{\mathrm{rank}}
\def \pf{{\bf Proof: }}
\def \dist{\mathrm{dist}}
\def \Re{\mathsf{Re}}
\def \Im{\mathsf{Im}}
\def \re{\mathsf{re}}
\def \im{\mathsf{im}}

\def \sym{\mathsf{sym}}
\def \sksym{\mathsf{skew\mbox{-}sym}}
\def \r{\mathrm{r}}
\def \even{\mathrm{even}}
\def \herm{\mathsf{Herm}}
\def \skherm{\mathsf{skew\mbox{-}Herm}}
\def \str{\mathrm{ Struct}}
\def \eproof{$\blacksquare$}
\def \proof{\noin\pf}

\def \bS{{\bf S}}
\def \cA{{\cal A}}
\def \E{{\mathcal E}}
\def \X{{\mathcal X}}
\def \F{{\mathcal F}}
\def \tr{\mathrm{Tr}}
\def \range{\mathrm{Range}}

\def \pal{\mathrm{palindromic}}
\def \palpen{\mathrm{palindromic~~ pencil}}
\def \palpoly{\mathrm{palindromic~~ polynomial}}
\def \hodd{H\mbox{-}\odd}
\def \heven{H\mbox{-}\even}
%\def \herm{\mathrm{Hermitian}}
%\def \skherm{\mathrm{skew\mbox{-}Hermitian}}
%\def \str{\mathrm{ Struct}}

\title{Context dependent preferential attachment model for complex networks}
\author{ Pradumn~Kumar~Pandey
		and~Bibhas~Adhikari
\thanks{Centre for System Science,
IIT Jodhpur, India, E-mail: pg201283006@iitj.ac.in}
\thanks{Centre for System Science,
IIT Jodhpur, India, E-mail: bibhas@iitj.ac.in}}

\date{}

\maketitle \thispagestyle{empty}

\maketitle

\begin{abstract}
 In this paper, we propose a growing random complex network model, which we call context dependent preferential attachment model (CDPAM), when the preference of a new node to get attached to old nodes is determined by the local and global property of the old nodes. We consider that local and global properties of a node as the degree and relative average degree of the node respectively. We prove that the degree distribution of complex networks generated by CDPAM follow power law with
exponent lies in the interval [2, 3] and the expected diameter grows logarithmically with the size of new nodes added in the initial small network. Numerical results show that the expected diameter stabilizes when alike weights to the local and global properties are assigned by the new nodes. Computing various measures including clustering coefficient, assortativity, number of triangles, algebraic connectivity, spectral radius, we show that the proposed model replicates properties of real networks better than BA model for all these measures when alike weights are given to local and global property. Finally, we observe that the BA model is a limiting case of CDPAM when new nodes tend to give large weight to the local property compared to the weight given to the global property during link formation.

\end{abstract}

\begin{IEEEkeywords}
context dependent preferential attachment, degree, relative average degree, clustering coefficient, assortativity, number of triangles, algebraic connectivity, spectral radius, diameter.
\end{IEEEkeywords}

% \fi
\IEEEpeerreviewmaketitle

\section{Introduction}

Modelling complex networks has been an active area of research in literature due to its applications in various field of science and technology \cite{murray2002innovation}\cite{wang2005general}\cite{balthrop2004technological}\cite{bullmore2009complex}. Several attempts have been made to generate deterministic and random complex network models which can capture the spirit of several large scale real world networks such as social networks \cite{Cd13}, biological networks \cite{maere2005bingo}, technological networks \cite{albert2004structural}etc. Two prime characteristics of a large class of real networks that have been observed and established by leading scholars in the area of complex networks are power-law degree distribution of the nodes and small-world behavior of the networks \cite{Cd28} \cite{Cd42}\cite{Cd37}\cite{BarabasiAlb99}\cite{Cd9}. The Erd$\ddot{o}$s-R$\acute{e}$nyi (ER) model \cite{erdds1959random} is one of the first initiatives to generate random networks where the links are made by following a random procedure when a fixed number of nodes is chosen at the initial stage of the network formation. However, later it has been observed that ER model fails to represent the essence of real networks, for example, degree distribution is not a power-law.  Consequently, a lot of interest has been generated to produce networks having power-law degree distributions.

One of the insightful growing random complex network models is proposed by Barabasi et al. in 1999, also called BA-model \cite{BarabasiAlb99}. In this model, a small network is chosen in the beginning of the method, then new nodes appear and get linked with the existing nodes in a probabilistic fashion which is decided by the property (degree) of the existing nodes \cite{BarabasiAlb99,AlbjeBarabasi99}. The philosophy adopted here is that at each iteration, the new nodes prefer to get attached with an old node which has high degree (among all existing ones) which sometimes represent the importance of a node in social context. Interestingly, the network generated by this model has power-law degree distribution and thus the concept of scale-free networks emerged. In his seminal paper \cite{BarabasiAlb99}, Barabasi et al. have also predicted that the growth and preferential attachment are jointly responsible for the emergence of the scale-free property in real networks. It has also been shown that the diameter grows approximately logarithmically with the size of the network.

Does a new node always wish to form links with important (high degree) nodes or the choice get influenced by other factors also? Moreover, if the choice gets influenced by other properties of the existing nodes, will the network be having power-law degree distribution? An evidence of a phenomena that peoples choice does not depend on only one property is given in \cite{Cd38}  supported by an empirical data (see \cite{Cd39}\cite{Cd40}\cite{Cd41} also). The data shows that at the time of purchasing a product, a buyer considers the background (history) of the product and relative attractiveness of the product with respect to other products in the same reference. Thus, the concept of context preferential attachment was introduced in \cite{Cd38}.

In this paper, we propose a growing random complex network model where the probability of link formation is determined by weighted local and global property of the existing nodes. We consider that local and global properties of a node are given by the degree and relative average degree of the node in a network. Thus, we call the proposed model, the context dependent preferential attachment model (CDPAM) for complex networks. We prove that the degree distribution of complex networks generated by CDPAM follow power law $P(k)=L(k)k^{-\gamma}$ where $2\leq \gamma \leq 3$ and $L(k)\rightarrow \alpha$ (a constant which depends on the weights given on local and global property of the nodes) as $k\rightarrow \infty.$ We also prove that the expected diameter grows logarithmically with the size of the new nodes added in the network, however the growth of the expected diameter is slower than that of the BA model. However, our numerical simulations show that the expected diameter stabilizes when alike weights are given to the local and global property which determine the preference of link formation. In contrast to the conventional wisdom that diameter shows as a function of $\ln(\ln N)$ or $\ln N$ in real networks, the authors in \cite{leskovec2007graph} observed that the diameter stabilizes or shrinks as a network grows. The proposed model reveals how shrinking and increasing of diameter are related to the weights on local and global property of the nodes during expansion of the network.    

A variety of mathematical and statistical measures have been proposed in the literature in order to characterize global and local structure of complex networks. We derived clustering coefficient, assortativity, number of triangles, algebraic connectivity, spectral radius for different complex networks generated by CDPAM and compare them with the same obtained from the complex network generated by BA model. We show that our model replicates properties of real networks better than BA model for all these measures when alike weights are given to local and global property. Finally, we observe that the BA model is a limiting case of CDPAM when new nodes tend to give large weight to the local property compared to the weight given on the global property during link formation.

\section{Context dependent preferential attachment model (CDPAM)}

In this section, we propose a random complex network model which relies on the fact that the network is open i.e. a network continuously grows in time with the addition of new nodes in to a fixed small network chosen in the beginning of the process \cite{BarabasiAlbJeo99}. It is important to notice that the link formation in BA model is biased as the link formation depends only on the high degree (importance) of the existing nodes. However, in real life we prefer to form relationship (link) with important (global property) people in society but also give importance to background (local property) of the people before making the relation. Inspired by this thought, we introduce the model as follows. \begin{enumerate} \item Growth: Starting with a small network having $m_0$ nodes, at every timestep we add a new node with $m \leq m_0$ edges and the new nodes get linked with the nodes already present in the network. \item Context preferential attachment: Assume that $N(t)$ denotes the node set of the network after $t$-time step. When a new node $j$ appears at time $t+1$ would get connected to node $i\in N(t)$ with probability $p_j^i(t+1)$ given by \be p_j^i(t+1)=\frac{\beta \, f_B(i) + \theta \, g(i,N(t))}{\sum_{i\in N(t)}(\beta \, f_B(i) + \theta \, g(i,N(t)))}\ee where $f_B(i)$  quantifies the background (local context) of node $i$, $g(i,N(t))$ determines the relative advantage (global context) of a nodes over others in the network $N(t),$ and $\beta, \theta (<\beta)$ are the positive control parameters for the property of the nodes in $N(t)$.
\end{enumerate}

In order to simplify the model, we consider $$f_B(i)=k_i \, \mbox{and} \, g(i,N(t))=\frac{\sum_{l\in N(t)} k_i-k_l}{|N(t)|}$$ where $k_i$ denotes the degree of a node $i$ and $|N(t)|$ is the number of nodes in $N(t).$ As we consider that a single node appears at each timestep, after time t there will be $t+m_0$ nodes in the network and for a large value of $t (\gg m_0),$ $|N(t)|\approx t.$ Consequently, we have \begin{equation}
\begin{split} 
p_j^i(t+1) &\approx  \frac{\beta k_i + \theta \sum_{l\in N(t)} \frac{k_i-k_l}{t}}{\sum_{l\in N(t)} \beta k_l + \sum_{l\in N(t)} \frac{(t+m_0)k_l - 2mt-m_0(m_0-1)}{t}} \\
&\approx  \frac{\beta k_i + \theta(k_i-2m)}{2m\beta t}
\nonumber
\end{split}
\end{equation} 
for a very small value of $m_0.$ Assuming $k_i$ to be a continuous real variable function and the rate of change of $k_i$ is proportional to $p_i^j(t),$ we have \be\label{eqn:p} \frac{\partial k_i}{\partial t} = m\frac{\beta k_i + \theta (k_i-2m)}{2m\beta t}\ee by applying mean field theory.

The degree distribution of the network generated by the CDPAM is provided in the following theorem.

\begin{theorem}
\label{th1}
The degree distribution of a complex network generated by CDPAM described above exhibits a power law in their tail given by $P(k)=L(k)k^{-\gamma}$ where $L(k)\rightarrow (\gamma -1) (m-c)^{(\gamma -1)}$ as $k\rightarrow \infty$ and $\gamma=1+\frac{2\beta}{\beta+\theta}, c=\frac{2m\theta}{\beta+\theta}.$ In particular, $\gamma\approx 2$ if $\beta\approx \theta$ and $\gamma\approx 3$ if $\beta\gg \theta.$
\end{theorem}
\pf From (\ref{eqn:p}) we have $$\frac{\partial k_i}{\partial t} = m\frac{\beta k_i + \theta (k_i-2m)}{2m\beta t} = \frac{k_i-c}{(\gamma -1)t}$$ solving which we obtain \be\label{eqn:deg}k_i(t)=(m-c)\left(\frac{t}{t_i}\right)^{1/(\gamma -1)}+c\ee when the initial condition is given by $k_i(t_0)=m.$ This yields $$P(k_i(t)< k)= P(t_i > (m-c)^{\gamma-1}(k-c)^{1-\gamma}t).$$ Assuming $k_i(t)< k,$ we have $t_i > (m-c)^{\gamma -1}(k-c)^{1-\gamma}t.$ Further, since it is assumed that a single node gets added at each timestep, it is equivalent to a uniform distribution of $t_i$, given by $P(t_i)=1/(m_0+t).$ Consequently, \beano P(k_i(t)< k) &= P(t_i> (m-c)^{\gamma-1}(k-c)^{1-\gamma}t)\\&=1-\frac{t}{t+m_0}(m-c)^{\gamma-1}(k-c)^{1-\gamma}\eeano The degree distribution is obtained by $$P(k)=\frac{\partial P(k_i(t)<k)}{\partial k} = \frac{t}{t+m_0}(\gamma -1)(m-c)^{\gamma -1}(k-c)^{-\gamma}.$$ The desired result for degree distribution follows from the fact that $t\rightarrow \infty.$

\begin{figure}
\begin{subfigure}{.25\textwidth}
   \includegraphics[width=4.90cm]{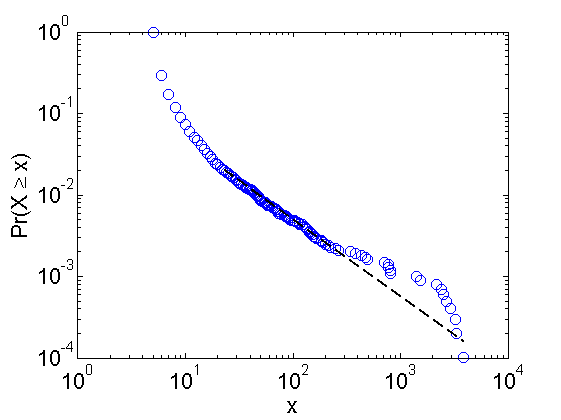}
  \caption{\footnotesize $\beta=0.6$, $\theta=0.5$}
  \label{fig:sub11}
\end{subfigure}%
\begin{subfigure}{.25\textwidth}
   \includegraphics[width=4.90cm]{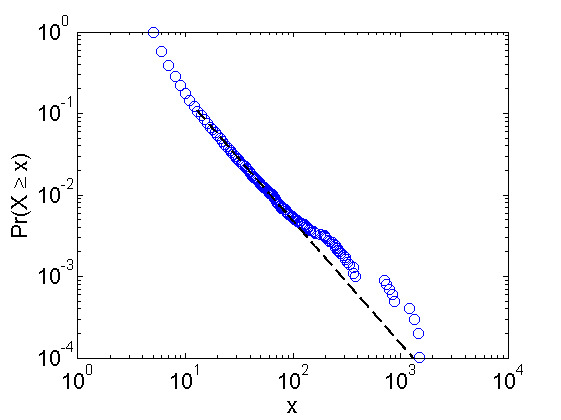}
  \caption{\footnotesize $\beta=1.2$, $\theta=0.5$}
  \label{fig:sub12}
\end{subfigure}
\begin{subfigure}{.25\textwidth}
   \includegraphics[width=4.0cm]{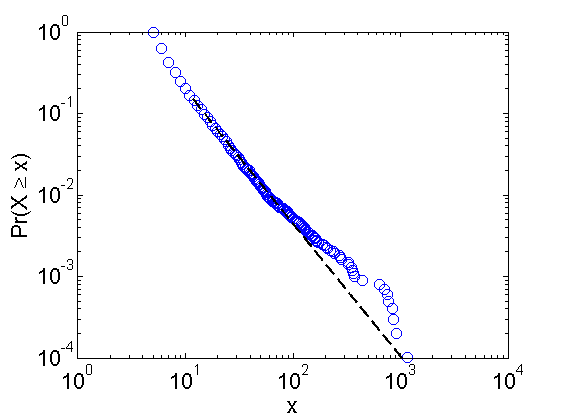}
  \caption{\footnotesize $\beta=1.8$, $\theta=0.5$}
  \label{fig:sub13}
\end{subfigure}%
\begin{subfigure}{.25\textwidth}
   \includegraphics[width=4.90cm]{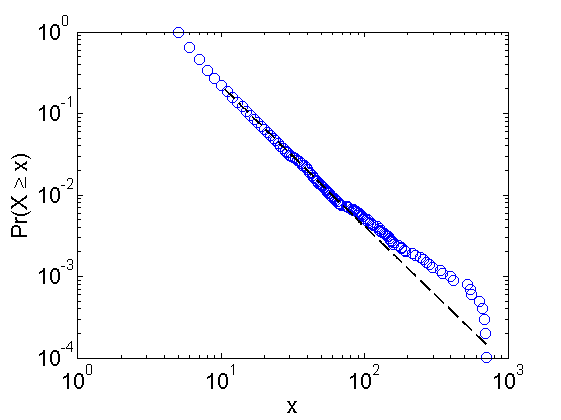}
  \caption{\footnotesize $\beta=2.4$, $\theta=0.5$}
  \label{fig:sub14}
\end{subfigure}
\begin{subfigure}{.25\textwidth}
   \includegraphics[width=4.90cm]{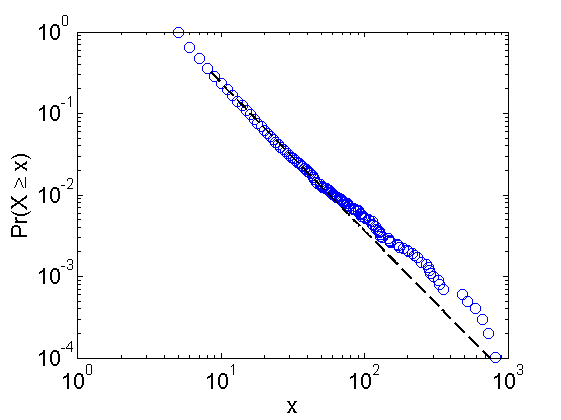}
  \caption{\footnotesize $\beta=3.0$, $\theta=0.5$}
  \label{fig:sub15}
\end{subfigure}%
\begin{subfigure}{.25\textwidth}
   \includegraphics[width=4.90cm]{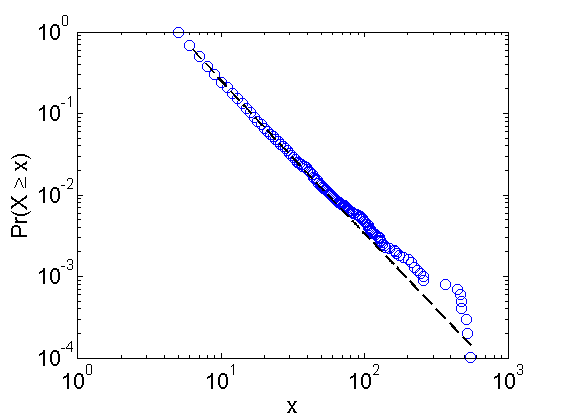}
  \caption{\footnotesize $\beta=6$, $\theta=0.5$}
  \label{fig:sub16}
\end{subfigure}
\begin{subfigure}{.25\textwidth}
   \includegraphics[width=4.90cm]{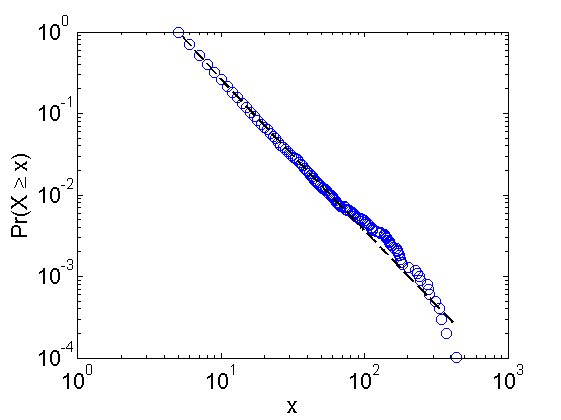}
  \caption{\footnotesize $\beta=60$, $\theta=0.5$}
  \label{fig:sub17}
\end{subfigure}%
\begin{subfigure}{.25\textwidth}
   \includegraphics[width=4.90cm]{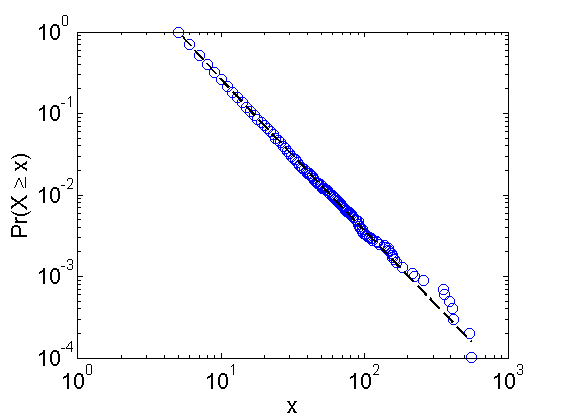}
  \caption{\footnotesize $\beta=300$, $\theta=0.5$}
  \label{fig:sub18}
\end{subfigure}
\begin{subfigure}{.25\textwidth}
   \includegraphics[width=4.90cm]{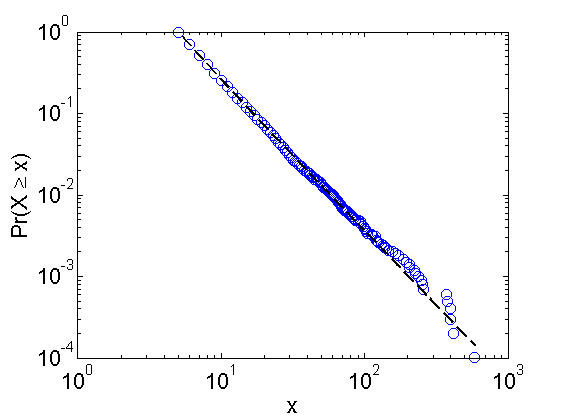}
  \caption{\footnotesize $\beta=600$, $\theta=0.5$}
  \label{fig:sub19}
\end{subfigure}%
\begin{subfigure}{.25\textwidth}
   \includegraphics[width=4.90cm]{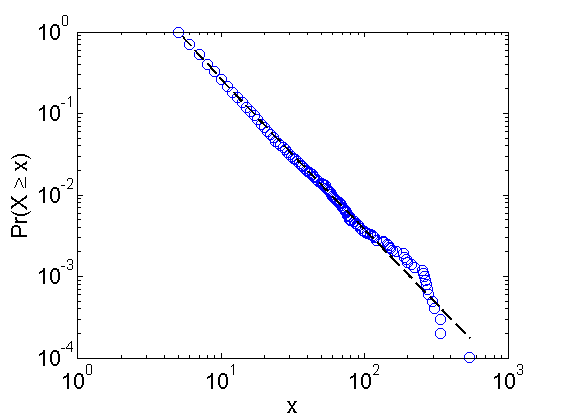}
  \caption{\footnotesize $\beta=600000$, $\theta=0.5$}
  \label{fig:sub110}
\end{subfigure}
\caption{ Degree distribution % (vertical axis represents information level and horizontal axis represents time steps).\ref{fig:sub1}  Watts- Strogatz model \cite{Cd19}, \ref{fig:sub2} Albert-Barabasi model \cite{Cd21}\cite{Cd22},\ref{fig:sub3}  ER model \cite{Cd20},\ref{fig:sub4} hierarchical model \cite{Cd23}\cite{Cd24}\cite{Cd25}, \ref{fig:sub6} Collaboration network of Arxiv General Relativity, \ref{fig:sub7} ego-Facebook Network, \ref{fig:sub8} Collaboration network of Arxiv High Energy Physics Theory
}
\label{fig 3}
\end{figure}

Setting the initial network the complete network with $7$ nodes, i.e. $m_0=7$ and $m=5$, we plot degree distributions of complex networks generated by CDPAM for different values of $\beta$ and $\gamma$ in Fig \ref{fig 3}. We also calculate the $p$-value which is a measure of goodness-of-fit based on KS statistics, to validate the power-law degree distribution of the networks \cite{Cd42}. The numerical simulations show that the exponent $\gamma$ is an increasing function of $\beta$ when $\theta$ is fixed.

\begin{table}[h!]

\begin{center}

\begin{tabular}{|c|c|c|c|} \hline\hline

\emph{\footnotesize  $\beta$, $\theta=0.5$}&\emph{\footnotesize $\gamma$ (calculated Numerically)} & \emph{\footnotesize p-value } & \emph{ \footnotesize $\gamma$ (Theoretical)}
\\

\hline
0.6	&1.94	&	0.170	&	2.090	\\
1.2	&2.50	&	0.090	&	2.411	\\
1.8	&2.62	&	0.220	&	2.565	\\
2.4	&2.70	&	0.490	&	2.655	\\
3.0	&2.78	&	0.135	&	2.714	\\
6	&2.84	&	0.025	&	2.846	\\
60	&2.82	&	0.600	&	2.980	\\
300	&2.82	&	0.996	&	2.996	\\
600	&2.82	&	0.290	&	2.998  \\
600000	&2.81	&	0.017	&	2.999  \\

\hline\hline

\end{tabular}

\end{center}

\caption{ Network parameters} \label{table1}

\end{table}

In order to show that the diameter of complex network constructed by the CDPAM is small, we proceed as follows. Let the node $i$ and $j$ appeared in the network at time $t_i$ and $t_j$ respectively. Assume that $t_i< t_j.$ Then the probability of the node $j$ to be linked with the node $i$ is given by $$p_j^i = m\frac{\beta k_i(t_j) + \theta (k_i(t_j)-2m)}{2m\beta t_j}$$ where $k_i(t_j) = (m-c)\left(\frac{t_j}{t_j}\right)^{1/(\gamma -1)}+c$ (see (\ref{eqn:deg})) is the degree of the node $i$ at time $t_j.$ Thus, \be\label{eqn:pij}p_j^i = \frac{m-c}{(\gamma -1)t_i^{1/(\gamma -1)}t_j^{1-1/(\gamma -1)}} + \frac{m(1-2\theta)}{2\beta t_j}.\ee

\begin{remark}
It is evident from the above derivation that the control parameters $\beta$ and $\theta$ which represent weights to the local and global property of the existing nodes respectively, determine the topology of the network generated by CDPAM. A natural question would be: Does there exist a functional relation between these parameters? To investigate how different values of these parameters affect the topology of the network, we fix the parameter $\theta$ and vary $\beta$ in the sequel. Thus, now onward we set $\theta=0.5.$
\end{remark}

%
%\begin{figure}
%\begin{subfigure}{.25\textwidth}
%   \includegraphics[width=4.90cm]{A_1000.png}
%  \caption{\footnotesize $\beta=600$, $\theta=0.5$}
%  \label{fig:sub19}
%\end{subfigure}%
%\begin{subfigure}{.25\textwidth}
%   \includegraphics[width=4.90cm]{A_1000000.png}
%  \caption{\footnotesize $\beta=600000$, $\theta=0.5$}
%  \label{fig:sub110}
%\end{subfigure}
%\caption{ Degree distribution % (vertical axis represents information level and horizontal axis represents time steps).\ref{fig:sub1}  Watts- Strogatz model \cite{Cd19}, \ref{fig:sub2} Albert-Barabasi model \cite{Cd21}\cite{Cd22},\ref{fig:sub3}  ER model \cite{Cd20},\ref{fig:sub4} hierarchical model \cite{Cd23}\cite{Cd24}\cite{Cd25}, \ref{fig:sub6} Collaboration network of Arxiv General Relativity, \ref{fig:sub7} ego-Facebook Network, \ref{fig:sub8} Collaboration network of Arxiv High Energy Physics Theory
%}
%\label{fig 3_1}
%\end{figure}

We recall the following lemma from \cite{Cd8}.

\begin{lemma}
If $A_1, A_2, ... A_n$ are mutually independent events and their probabilities full fill the relations $P(A_i)\leq \epsilon$ for all $i$ then
\begin{equation}
P \left( \bigcup_{i=1}^n A_i\right)=1-exp\left( - \sum_{i=1}^n P(A_i)\right)-Q
\nonumber
\end{equation}
where $0 \leq Q < \sum_{j=0}^{n+1}(n \epsilon)^j/j!-(1+\epsilon)^n.$
\end{lemma}

Assume that $N(t)$ denotes the set of all nodes which have been added in the network up to timestep $t.$  In the network generated by CDPAM, assume that the nodes $i,j\in N(t)$ are connected by a path $(i, v_1, v_2, \hdots, v_{l-1},j)$ of length $l$ where $v_k\in N(t)$ for all $k=1:l-1.$ Consider that this sequence is a single event $A_k.$ The total number of such events possible is $|N(t)|^{l-1}.$ Thus, as given in \cite{Cd8}, the probability of the existence of a path between $i$ and $j$ of length not more than $l$ is given by
 \begin{equation}
 \label{eqn:pl} 
 \begin{split}
 P_{ij}(l) &= P\left(\bigcup_{k=1}^{|N(t)|^{l-1}} A_k\right)\\
 &=1-exp\left[- \sum_{v_1=1}^{|N(t)|} \hdots \sum_{v_{l-1}=1}^{|N(t)|} p_{i}^{v_1} \hdots p_{v_{l-1}}^j \right]
  \end{split}
 \end{equation}
 
  We use this result to obtain the following corollary.

\begin{corollary}
The probability of the existence of a path between two vertices $i,j\in N(t)$ of length not more than $l$ is given by $$P_{ij}(l)=1-exp\left[- \dfrac{K^{l}H_{n}^{l-1}}{t_i^{1/(\gamma-1)}t_j^{1-1/(\gamma-1)}}  \right]$$ where $K=\frac{(\beta+0.5)(m-c)}{2\beta}, H_n=\sum_{k=1}^{|N(t)|}\frac{1}{k}$ and $c$ is given in theorem \ref{th1}.
\end{corollary}
\pf Using (\ref{eqn:pij}) and (\ref{eqn:pl}) the result follows.

 \begin{corollary}\label{Cor:2}
 The expected value $l_{ij}$ of the distance between two nodes $i,j\in N(t)$ is given by $$l_{ij}=\dfrac{\left(1-\dfrac{1}{\gamma-1}\right)\ln t_j+\dfrac{1}{\gamma-1}\ln t_i-\log K-r}{\ln (K H_n)}+\dfrac{1}{2}.$$
 \end{corollary}
 \pf The result follows from the fact that $$l_{ij}=\sum_{l=0}^{\infty} F(l)$$ where $F(l)=1-P_{ij}(l)$ (see \cite{Cd8}).

 Observe in Corollary \ref{Cor:2} that the expected distance  $l_{ij}$ between two nodes $i,j\in N(t)$ is an increasing function of $t_i$ and $t_j$ when other parameters are fixed. This implies that the diameter of the network is the expected distance between the first node and the last node added in the network. Hence, setting $t_j=|N(t)|$ and $t_i=1$ we obtain the following result.

 \begin{corollary}
 The expected diameter of a complex network generated by CDPAM is given by $$D=\dfrac{\left(1-\dfrac{1}{\gamma-1}\right)\ln |N(t)|-\ln K -r}{\ln (K H_n)}+\dfrac{1}{2}.$$
 \end{corollary}

\begin{figure}
  \begin{center}
   \includegraphics[width=7.0cm]{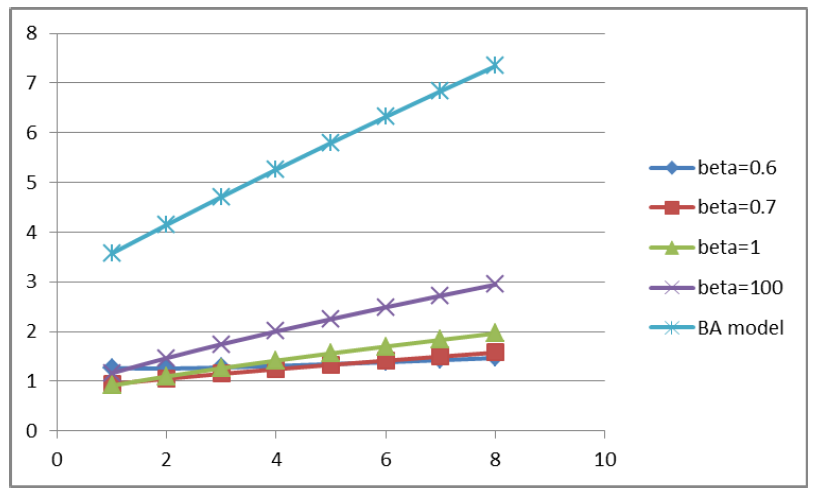}
  \caption{\footnotesize Diameter growth of networks Horizontal-axis represents the $\ln N$ and Vertical-axis represents $D$ }
\label{fig 8}
\end{center}
\end{figure}

 Thus it follows from the above corollary that the expected diameter of the network depends on the logarithmic value of the size of new nodes added in the  network.
 In Fig.\ref{fig 8}, we calculated the expected diameter for CDPAM and the approximate diameter given by BA model $(\sim \ln N/ \ln \ln N)$ \cite{PhysRevLett.90.058701}. However, numerical simulations show that the expected diameter of CDPAM stabilizes when alike weights are assigned to both the local and global properties which determine the preference of link formation. In contrast to the conventional wisdom that diameter is a function of $\ln(\ln N)$ or $\ln N$ in real networks, the authors in \cite{leskovec2007graph} observed that the diameter stabilizes or shrinks as a network grows. The CDPAM  reveals how shrinking and increasing of diameter are related to the weights on local and global property of the nodes during expansion of the network.

 \section{Property of complex networks generated by CDPAM}

 In this section, we numerically calculate various measures which include clustering coefficient, assortativity, algebraic connectivity, and spectral radius for the complex networks generated by CDPAM. These measures determine various topological features of the network and enable to compare how the proposed model captures the property of different real networks. We also compare values of these measures with that of complex network generated by BA model. We have used MATLAB R2012a for the numerical simulations.

 \subsection{Clustering coefficient}
\begin{figure}
 \begin{center}
   \includegraphics[width=7.0cm]{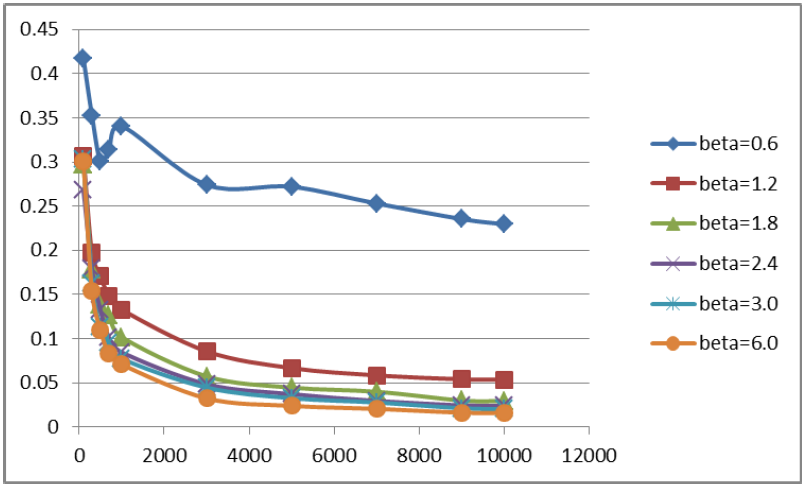}
  \caption{\footnotesize Horizontal-axis represents size of the network and Vertical-axis represents clustering coefficient}
\label{fig 1}
\end{center}
\end{figure}

\begin{figure}
 \begin{center}
   \includegraphics[width=7.0cm]{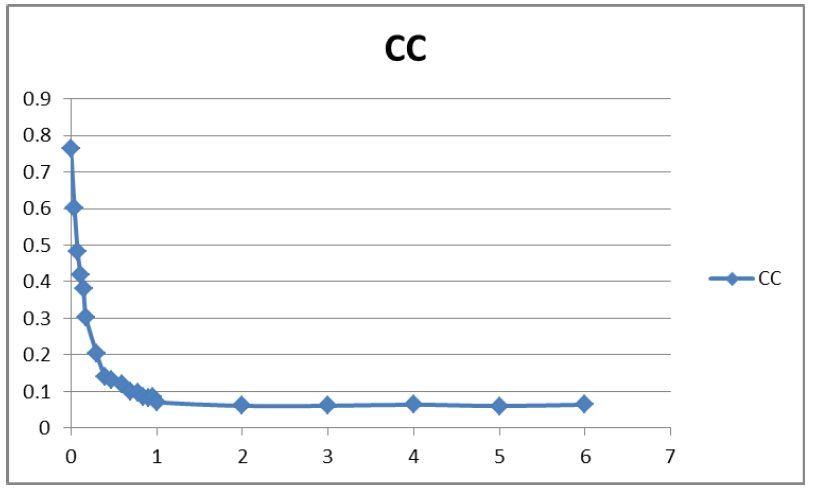}
  \caption{\footnotesize Horizontal-axis represents log(beta) and Vertical-axis represents clustering coefficient of the network }
\label{fig 2}
\end{center}
\end{figure}

 Clustering coefficient (CC) of a node signifies the local edge density among the neighbors of the node. The CC of a network is the average of CC of all the nodes. Thus, for a network $N,$ $$CC(i)=\frac{2|E_i|}{k_i(k_i-1)} \,\, \mbox{and} \,\, CC(N)=\frac{1}{|N|}\sum_i CC(i)$$ where $|E_i|$ denotes the number of links adjacent to a node $i$ of a network \cite{Cd28}. It is evident that $0\leq CC(N)\leq 1$ for any network $N.$ In Fig. \ref{fig 1}, we plot the CC of different size of complex networks generated by CDPAM with different values of $\beta$ and $\theta=0.5.$ It shows that as the value of $\beta$ increases the CC of the network decreases and eventually when $\beta$ is very large, the CC is close to the CC of the network generated by BA model. The Fig \ref{fig 2} shows that the CC gets close to $0.8$ as $\log \beta$ gets close to zero. Thus, we conclude that, in CDPAM model, if link is formed by giving equal weights to local and global properties of the existing node then the CC gets close to $0.8$ which is a property of a large class of real networks like ego-Facebook network, ego-Gplus network, ego-Twitter \cite{Cd13}.
 
 \subsection{Assortativity index}
 The Assortative Index (AI) of a network $N$ is defined by $$AI(N)=\frac{\sum_{ij} (a_{ij}-\frac{k_ik_j}{2m})k_ik_k}{\sum_{ij} (k_i\delta_{ij}-\frac{k_ik_j}{2m})k_ik_j}$$ where $a_{ij}$ is the $ij$-th entry of the adjacency matrix associated with $N,$ $\delta_{ij}$ is the Kronecker delta function  \cite{Cd31}. Obviously $-1\leq AI(N)\leq 1.$ A positive value of $AI(N)$ signifies nodes with similar degree nodes are linked whereas a negative value of $AI(N)$ implies that similar degree nodes are not linked. 
 
\begin{figure}
  \begin{center}
   \includegraphics[width=7.0cm]{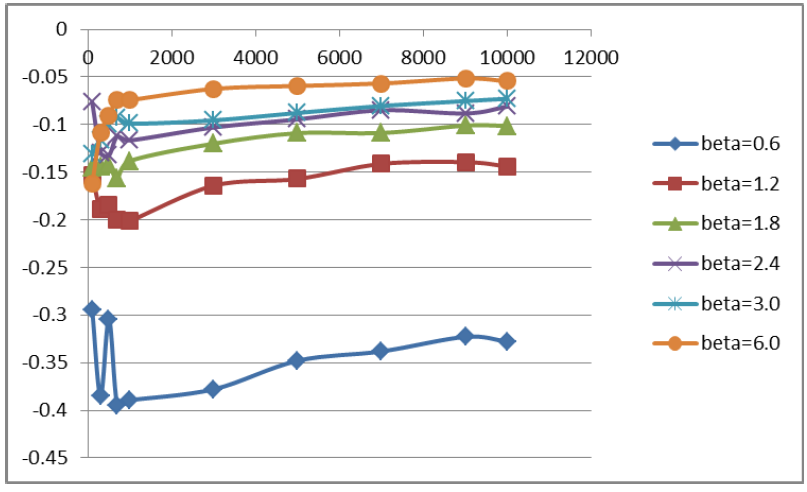}
  \caption{\footnotesize  Horizontal-axis represents size of the network and Vertical-axis represents assortativity of the network}
\label{fig 4}
\end{center}
\end{figure}

 Consider the network $N$ after addition of $t$ nodes to the given small network.Then by (\ref{eqn:deg}), it follows that degree of a not is a decreasing function of timestep of its appearance. Further, if a node $j$ which appeared in the network at the $t_j$ timestep has probability $p^i_j$ to get linked with an existing node $i$ appeared at $t_i<t_j$, is a decreasing function in both $t_i$ and $t_j$, see (\ref{eqn:pij}). These indicate, the probability of having a link between high degree nodes is larger compared to the probability of having a link in between low degree nodes. Therefore, we conclude that the network is assortative for higher degree nodes and disassortative for low degree nodes. Since the network has a few high degree nodes, overall the network is disassortative. The plots given in Fig \ref{fig 4} assert the same for different values of $\beta$ and $\theta=0.5.$ We mention here disassortative phenomena of networks occur in a large class of real networks including World-Wide-Web \cite{BarabasiAlb99}, Marine food web \cite{huxham1996parasites}, freshwater food web \cite{martinez1991artifacts}.
 
\subsection{Number of triangles}
\begin{figure}
\begin{center}
   \includegraphics[width=7.0cm]{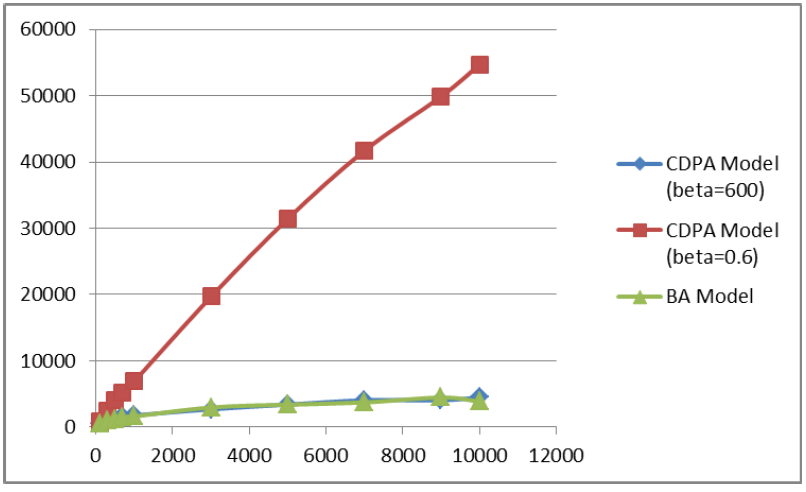}
  \caption{\footnotesize Horizontal-axis represents size of the network and Vertical-axis represents triangle count of the network}
\label{fig 5}
\end{center}
\end{figure}

A triangle is a cycle with three nodes. The number of triangles is a fundamental building block for many real networks. In a social network, if nodes are human beings and links are described by friendship relation, then the a triangle means friends of a friend are friends. Often real networks consists of a huge number of triangles which could be both homogeneous and heterogeneous \cite{durak2012degree}. In Fig \ref{fig 5}, we show that the proposed complex networks by CDPAM contain huge number of triangles compared to a network constructed by the BA model for example ego-Facebook network, ego-Gplus network, ego-Twitter \cite{Cd13}.

\subsection{Algebraic connectivity}
\begin{figure}
\begin{center}
   \includegraphics[width=7.0cm]{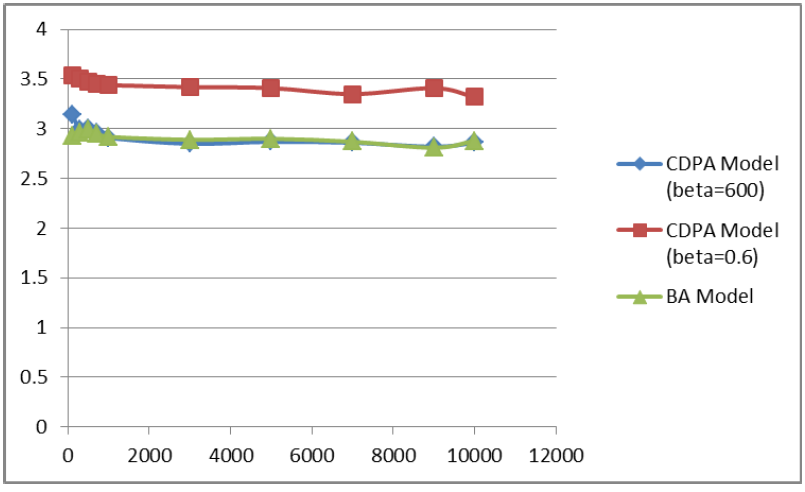}
  \caption{\footnotesize  Horizontal-axis represents size of the network and Vertical-axis represents algebraic connectivity of the network}
\label{fig 6}
\end{center}
\end{figure}

Algebraic connectivity of a network $N$ is the second largest eigenvalue of the Laplacian matrix $L(N)=D(N)-A(N)$ associated with the network where $D(N)=\diag\{k_1, \hdots, k_n\}$ denotes the degree matrix and $A(N)$ is the adjacency matrix of the network \cite{Cd33}. Obviously, $L(N)$ is a symmetric positive semi-definite matrix. It is well known that the second eigenvalue $\lambda_2$ of $L(N)$ is positive if and only if $N$ is connected. More importantly, $\lambda_2$ determines the robustness of a network, i.e. larger the value of $\lambda_2,$ the more difficult to make the network disconnected by removal of nodes or edges \cite{Cd33}. In particular, if $\mu(N)$ and $\eta(N)$ denote the vertex and edge connectivity of a network $N$ respectively, then $\lambda_2\leq \mu(N) \leq \eta(N).$ We show in Fig \ref{fig 6} that if a complex network is produced by CDPAM after setting $\beta\approx \theta,$ that is giving almost equal weighage to both local and global property of the existing nodes, then the network has higher algebraic connectivity than that of a network produced by the BA model.

\subsection{Spectral radius}
Spectral radius of a network is the maximum modulus of eigenvalues of the network. In \cite{Cd32} it has been shown that the reciprocal of the spectral radius decides the threshold of virus propagation in the network. The smaller the spectral radius is, the larger the robustness of a network against the spread of viruses \cite{Cd32}.
\begin{figure}
\begin{center}
   \includegraphics[width=7.0cm]{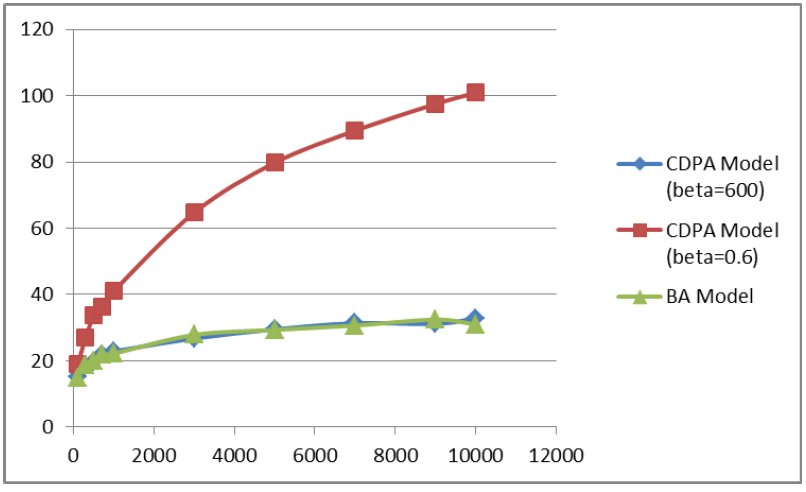}
  \caption{\footnotesize  Horizontal-axis represents size of the network and Vertical-axis represents spectral radius of the network}
\label{fig 7}
\end{center}
\end{figure}
\\
In Fig. \ref{fig 7} we plot the spectral radius of networks generated by CDPAM and compared with BA model.
 Real world networks show considerable larger spectral radius compared to BA model. CDPAM is capable to inherit large spectral radius as many real world networks including  Dutch soccer team network \cite{Cd32}, Dutch roadmap network \cite{jamakovic2006topological}, Internet graph at the IP-level \cite{jamakovic2006laplacian} and the Autonomous System level \cite{muhlbauer2006building}.  

\section{Conclusion}
In the literature of social choice theory and management science it has been established that the choice of a person get influenced by a given offered set and ultimately, the choice is determined by the local and global contexts of the items in the offered set. Inspired by this concept, we introduced a preferential attachment model for generating growing complex networks when the preference of a new node to get linked with old nodes in a network is determined by local and global properties of the old nodes. We call the model, the context dependent preferential attachment model (CDPAM) and the local property is given by the degree of a node, the global property is given by the relative average degree of the old nodes. We proved that the complex networks generated by CDPAM have power law degree distribution and expected diameter depends logarithmically with the size of new nodes added in the network. In contrast to the general intuition that diameter grows with the addition of new nodes, we numerically showed that, in the CDPAM model, the expected diameter stabilizes when the new nodes get linked by giving alike importance (weight) to both local and global property of the old nodes. 

In order to investigate how the complex networks generated by CDPAM and BA models are related, we calculated clustering coefficient, assortativity, number of triangles, algebraic connectivity, spectral radius for both the models. We compared these measures and concluded that BA model is a limiting case of CDPAM when new nodes tend to give large weight to the local property compared to the weight given to the global property during link formation. By using these measures, we showed that the CDPAM captures the properties of real networks better than BA model. 

An interesting question is: can communities emerge in CDPAM? We believe that communities will also emerge when the weights to the local and global properties will not be constant for all new nodes but vary with the new nodes. We plan to investigate this phenomenon in future.

% or


%\appendices
%% you can choose not to have a title for an appendix
%% if you want by leaving the argument blank
%\section{}
%% use section* for acknowledgement
%\section*{Acknowledgment}

%\ifCLASSOPTIONcaptionsoff
%  \newpage
%\fi



\bibliographystyle{IEEEtran}
\bibliography{science}
\end{document}